\documentstyle[aps,psfig,preprint]{revtex}
\begin{document}
\def\be{\begin{equation}}
\def\ee{\end{equation}}
\def\bfi{\begin{figure}}
\def\efi{\end{figure}}
\def\bea{\begin{eqnarray}}
\def\eea{\end{eqnarray}}
\title{Preasymptotic multiscaling in the phase-ordering dynamics
of the kinetic Ising model}

\author{Claudio Castellano\thanks{claudio@ictp.trieste.it}}
\address{The Abdus Salam International Centre for Theoretical Physics,
Strada Costiera 11, P.O. Box 586, 34100 Trieste, Italy}
\author{Marco Zannetti\thanks{zannetti@na.infn.it}}
\address{Istituto Nazionale di Fisica della Materia, Unit\`{a}
di Salerno and Dipartimento di Fisica, Universit\`{a} di Salerno,
I-84081 Baronissi (Salerno), Italy}
\maketitle

\begin{abstract}
The evolution of the structure factor is
studied during the phase-ordering dynamics of the kinetic Ising model
with conserved order parameter.
A preasymptotic multiscaling regime is found as in the solution
of the Cahn-Hilliard-Cook equation, revealing that the late stage
of phase-ordering is always approached through a crossover
from multiscaling to standard scaling,
independently from the nature of the microscopic dynamics.
\end{abstract}

\pacs{05.70.Fh, 64.60.Cn, 64.75.+g}

Recently we have made an extensive study\cite{Castellano98} of the overall
time evolution in the phase-ordering process with scalar conserved order
parameter in the framework of the Cahn-Hilliard-Cook (CHC)
equation\cite{Cahn58}.
Through the careful study of the global evolution of the structure
factor from the very beginning of the quench down to the fully developed
late stage, we have identified a pattern whereby the very early behavior
is followed by an intermediate nonlinear mean-field regime before the
asymptotic fully nonlinear regime is attained.
Furthermore, the time scales of these regimes have been found to be
strongly dependent on (i) the lengthscale considered and (ii) the parameters
of the quench such as the amplitude $\Delta$ of the initial fluctuations
and the final temperature $T_F$.
The concurrence of all these elements gives rise to a rich and interesting
variety of behaviors in the preasymptotic phenomenology which can be accounted
for in a simple manner on the basis of a model introduced by Bray and
Humayun (BH model)\cite{Bray92}.

In this paper we make an analogous study of the structure factor in the
framework of the kinetic Ising model evolving with Kawasaki dynamics.
The main point is that we observe also in this case
the existence of an intermediate regime
characterized by the mean-field multiscaling behavior, pointing to the
generic nature of the above described structure in the overall time evolution.

There is abundant experimental and numerical evidence for the universality
of the late stage scaling in the phase-ordering process\cite{Bray94}
yielding the following form for the structure factor
\be
C(\vec k,t) \sim L^d(t) F(kL),
\label{SS}
\ee
with
\be
L(t) \sim t^{1/z}.
\ee
In particular the dynamic exponent $z$ and the scaling function $F(x)$
are independent of the initial condition, of the final temperature
and of the nature of the dynamics, be it Langevin for continuous spins
or Kawasaki for discrete spins.
In a way this is easy to understand.
The behavior~(\ref{SS}) applies in the large time regime, when ordered
domains have formed and the process is dominated by the motion of
interfaces, obeying an effective equation of motion, which presumably
takes the same form independently from the nature of the microscopic
dynamics.
Now, with the results presented in this paper, we are able to extend
the independence from the nature
of the microscopic dynamics also to the preasymptotic
regime where mean-field behavior is observed.

In order to explain this let us summarize the results of
Ref.~\cite{Castellano98}.
The CHC equation of motion for an $N$-component order parameter
is given by
\be
{\partial {\vec \phi}({\vec x},t) \over \partial t} =
\nabla^2 
\left[ {\partial V({\vec \phi}) \over \partial {\vec \phi}} 
- \nabla^2{\vec \phi} \right]
+ {\vec \eta}({\vec x},t),
\label{4.1}
\ee
where $V({\vec \phi}) = (r/2) {\vec \phi}^2 + (g/4N)
({\vec \phi}^2)^2$, with $r<0$ and $g>0$, while
${\vec \eta}$ is a gaussian white noise, with expectations
\be
\left \{
\begin{array}{ccl}
<{\vec \eta}({\vec x},t)> & = & 0 \\
<\eta_{\alpha}({\vec x},t)\eta_{\beta}({\vec x}',t')> & =
& - 2 T_F  \delta_{\alpha \beta} \nabla^2
\delta({\vec x}-{\vec x}')\delta(t-t').
\end{array}
\right.
\ee
For future reference let us denote by $M_0^2= -r/g$ the square of the
order parameter at the bottom of the local potential, which we will take
$O(1)$.
Starting from this equation, through a combination of the $1/N$-expansion
and the gaussian auxiliary field approximation of Mazenko\cite{Mazenko89}, 
Bray and Humayun have derived a nonlinear closed equation for the structure
factor\cite{Bray92}
\be
{\partial C({\vec k},t) \over \partial t} = 
-2 k^2  [k^2+R(t)] C({\vec k},t) 
- 2{k^2 \over N} R(t) D({\vec k},t) + 2 k^2 T_F,
\label{BH}
\ee
where
\be
R(t) = r+g \int {d^dk \over (2\pi)^d} C({\vec k},t) 
\ee
and
\be
D({\vec k},t) = \int {d^dk_1 \over (2\pi)^d } \int {d^dk_2 \over (2\pi)^d }
C({\vec k}-{\vec k_1},t) C({\vec k_1}-{\vec k_2},t) C({\vec k_2},t).
\ee
Integrating formally with the initial condition
\be
C({\vec k},t=0)=\Delta
\ee
one can write the structure factor as the sum of three pieces
\be
C({\vec k},t) = \Delta C_0({\vec k},t) +
T_F C_{T_F}({\vec k},t) + {1 \over N} C_{nl}({\vec k},t)
\label{C_BH_tot}
\ee
where
\be
C_0({\vec k},t) = \exp{ \left \{-2 k^2\int_0^t dt' [k^2+R(t')]\right \}}
\ee
\be
C_{T_F}({\vec k},t) = 2 k^2 \int_0^t dt' {C_0({\vec k},t) \over
C_0({\vec k},t')}
\ee
\be
C_{nl}({\vec k},t) = -2 k^2 \int_0^t dt' {C_0({\vec k},t) \over
C_0({\vec k},t')} R(t') D({\vec k},t')
\ee
which are coupled together through the definitions of $R$ and $D$.

In general this is a very complicated nonlinear integral equation
which must be handled numerically\cite{Castellano96}.
However, as we have shown in Ref.~\cite{Castellano98},
progress in the understanding of the solution
can be made by considering the scaling properties of the three terms
separately.
Then, the first two terms, which are what is left after taking the
large-$N$ limit, obey multiscaling\cite{Coniglio89}.
Namely one has
\be
C_{0,T_F}({\vec k},t) \sim {\cal L}_1^{\alpha_{0,T_F}(x)} F_{0,T_F}(x)
\label{13}
\ee
with ${\cal L}_1=L(t)[k_m(t)L(t)]^{2/d-1}$, $L(t)=t^{1/4}$, $x=k{\cal L}_2$
and  ${\cal L}_2=k_m^{-1}(t)$ is the inverse of the peak wave vector.
The two lengths ${\cal L}_1$ and ${\cal L}_2$ are related by
\be
{{\cal L}_1 \over {\cal L}_2} = (d \ln L)^{1 \over 2d}.
\label{14}
\ee
The spectra of exponents $\alpha_{0,T_F}(x)$ are given by
\be
\alpha_0(x)=d \psi(x)
\ee
with
\be
\psi(x)= 1 - (1-x^2)^2
\ee
and
\be
\alpha_{T_F}(x) =
\left \{
\begin{array}{lcl}
2 + (d-2) \psi(x) & \hspace{1cm} & x < x^* \\
2                 & \hspace{1cm} & x > x^*
\end{array}
\label{MST}
\right.
\ee
where $x^*=\sqrt{2}$.
Finally $F_0(x)=1$ while $F_{T_F}(x)=T_F/x^2$.

The last term in the right hand side of~(\ref{C_BH_tot})
has been shown by Bray and Humayun\cite{Bray92} to obey standard scaling
\be
C_{nl}({\vec k},t) \sim L^d(t) F_{nl}(x).
\label{C}
\ee
Summarizing, the behavior of the full solution of~(\ref{BH}), is expected to
be the outcome of the competition of three terms of the type
\be
C({\vec k},t) \simeq \Delta L^{\alpha_0(x)} + T_F L^{\alpha_{T_F}(x)}
+ (1/N) L^d
\label{17}
\ee
where, for simplicity, we have omitted logarithmic factors taking
${\cal L}_1 = L$.

As it is evident from the plot of the spectra of exponents in Fig.~\ref{Fig1},
since $\alpha_{0,T_F}(x)\leq d$, for any finite $N$ the last term in the
right hand side of~(\ref{17}) eventually dominates on all lengthscales,
leading to the asymptotic standard scaling behavior of the structure factor.
However, due to the $x$ dependence of the exponents, the competition
takes place differently on different lengthscales.

For the full analysis of the interplay of the three terms we refer to
Ref.~\cite{Castellano98}.
In the present case with $d=2$, the second and third term in Eq.~(\ref{17})
scale in the same way and the interplay is only between the first
and the other two together.
Notice that the difference $\delta \alpha(x)= d-\alpha_0(x)$
vanishes for $x=1$, namely at the peak of the structure factor,
while it is positive for all $x \neq 1$ and in particular becomes
very large for $x>x^*$, where $\alpha_0(x)$ is negative.
Therefore, even if the last two terms are bound to dominate, by modulating
the choice of $\Delta$, the relative weight of the first term may be
adjusted so that multiscaling can be observed over a sizable
preasymptotic time interval, with a spread in $x$ about the peak
which depends on the actual value of $\Delta$.
With a continuous order parameter the value of $\Delta$ can be varied
at will and in Ref.~\cite{Castellano98}
it was shown that when $\Delta$ is very small
($\Delta \ll M_0^2$) there exist an observable mean-field preasymptotic
behavior practically only at the peak, while for large
values of $\Delta$ ($\Delta \simeq M_0^2$) preasymptotic mean-field
multiscaling is clearly observed over the range $x \lesssim x^*$.
The interest of this result is that it is not just a property of the BH model,
but the same pattern of behavior is observed in the simulation of the
full CHC equation\cite{Castellano98} with $N=1$.
We now show that a very similar structure in the overall time evolution
appears also in the dynamics of the Ising model.

The analysis of the crossover between multiscaling and standard scaling
is carried out by computing the effective spectrum of exponents $\alpha(x,t)$
that one obtains when fitting the structure factor to the form
\be
C(\vec{k},t) = {\cal L}_1^{\alpha(x,t)} F(x),
\label{8}
\ee
with $x=k{\cal L}_2$
and ${\cal L}_2(t)=k_m^{-1}(t)$. A delicate point here is the choice of
${\cal L}_1$. Adopting straightforwardly the definition of the large-$N$
model given after Eq.~(\ref{13}) amounts to put in by hand the growth law
$L(t)=t^{1/4}$. On the other hand, in the large-$N$ model ${\cal L}_1$
is also given by $[C_0(k_m,t)]^{1/d}$. Therefore we take as an
unbiased choice ${\cal L}_1=[C(k_m,t)]^{1/d}$, dropping an inessential
costant factor $F(1)$.
In practice $\alpha(x,t)$ is obtained as the slope of the plot of
$\ln C(x/{\cal L}_2(t),t)$ versus $\ln {\cal L}_1(t)$ for fixed $x$.
When the time dependence of $\alpha(x,t)$ disappears
there is scaling, which is of the
standard type if $\alpha(x)$ is independent of $x$ and it is of the
multiscaling type if $\alpha(x)$ actually displays a dependence on $x$.

We have considered an Ising system on a two dimensional lattice
of size 512$\times$512 with periodic boundary conditions and initially
prepared in an infinite temperature configuration.
We have then let the system evolve with Kawasaki spin-exchange
dynamics after a sudden
quench to a fixed temperature $T_F$ below the critical temperature $T_c$.
We have computed the structure factor by averaging over $10^2$
realizations of the time histories.
We have divided the entire
duration of the simulation ($10^5$ Monte Carlo steps/spin) into subsequent
and nonoverlapping time intervals, corresponding to temporal decades,
and computed the effective exponent in the different time intervals.

In Fig.~\ref{Fig2}
we have plotted the effective exponent as a function
of $x$ in the five decades for a quench to $T_F=0.5 T_c$.
The curves follow the pattern of the results
reported in Ref.~\cite{Castellano98} for the CHC equation when $\Delta$
is large, as indeed it is the case for the Ising system where $\Delta=1$.
The gross feature is that there are two markedly distinct behaviors
for $x<x^*$ and for $x>x^*$.
To the right of $x^*$ the curves display a clear time dependence
eventually reaching the standard scaling behavior $\alpha(x) \sim 2$
in the latest time interval. To the left of $x^*$, instead, the curves
are bunched together displaying multiscaling with a spectrum of
exponents which follows closely $\alpha_0(x)$.
This pattern fits quite well with the results of the previous analysis
of the competition between the first and the other two terms on the right
hand side of Eq.~(\ref{17}).
The multiscaling behavior for $x<x^*$ shows that the relative size of the
prefactors is such that, for these values of $x$, there exists a long
time interval during which the structure factor is dominated by $\Delta
L^{\alpha_0(x)}$.
This is the preasymptotic mean-field scaling which preceeds the crossover
towards the eventual standard scaling\cite{Amar88}.
The duration of our simulation allows to detect just the onset of the
crossover and it is not long enough to see the definitive establishment
of standard scaling.
For $x>x^*$, instead, $\delta \alpha(x)=d-\alpha_0(x)$ is much too big for
observing any multiscaling and the time dependence of the effective exponent,
jointly to the flat behavior as a function of $x$, shows that the crossover
towards standard scaling is fully under way.

The same analysis has been performed for a quench to $T_F = 0.9 T_c$,
obtaining results completely analogous to those of Fig.~\ref{Fig2}.
Moreover we have repeated the analysis for a one dimensional Ising model
quenched from infinite temperature to $T_F=0.5 J/k_B$.
For such value of $T_F$ the equilibrium correlation length is $\xi(T_F)
\simeq 27.3$ and the system orders for times such that the typical
domain length is smaller than $\xi(T_F)$. 
At the longest time in our simulation the typical domain size,
measured as the first zero of the correlation function, was $R_g=9.33$.
The behavior of
the effective exponent $\alpha(x,t)$ in Fig.~\ref{Fig3} is very
similar to the one in Fig.~\ref{Fig2}, displaying a neater approach
to standard scaling for $x>1$.

In summary,
the study of the overall time evolution in the kinetic Ising model evolving
with Kawasaki dynamics displays a clear crossover from multiscaling
to standard scaling as observed in the simulation of the CHC equation
with large initial fluctuations.
The first remark is that preasymptotic multiscaling is independent 
from the nature of the microscopic dynamics,
as well as asymptotic standard scaling.
The second remark is that multiscaling is the characteristic signature
of mean-field behavior.
Therefore the results presented in this paper together with those for the
CHC equation in Ref.~\cite{Castellano98}, show that a genuine feature of
spinodal decomposition is that before reaching asymptotics the
nonlinearity is governed by the mean-field mechanism on the large length
scales.
The reason of this phenomenon poses an interesting question for the theory
of phase-ordering kinetics. It would be also interesting to look
for systems with sufficiently large initial fluctuations
in order to observe multiscaling experimentally.

We thank the Istituto di Cibernetica of CNR (Arco Felice, Napoli) for 
generous grant of computer time. This work has been partially
supported from the European TMR Network-Fractals c.n. FMRXCT980183
and from MURST through PRIN-97.

\bfi
\centerline{\psfig{figure=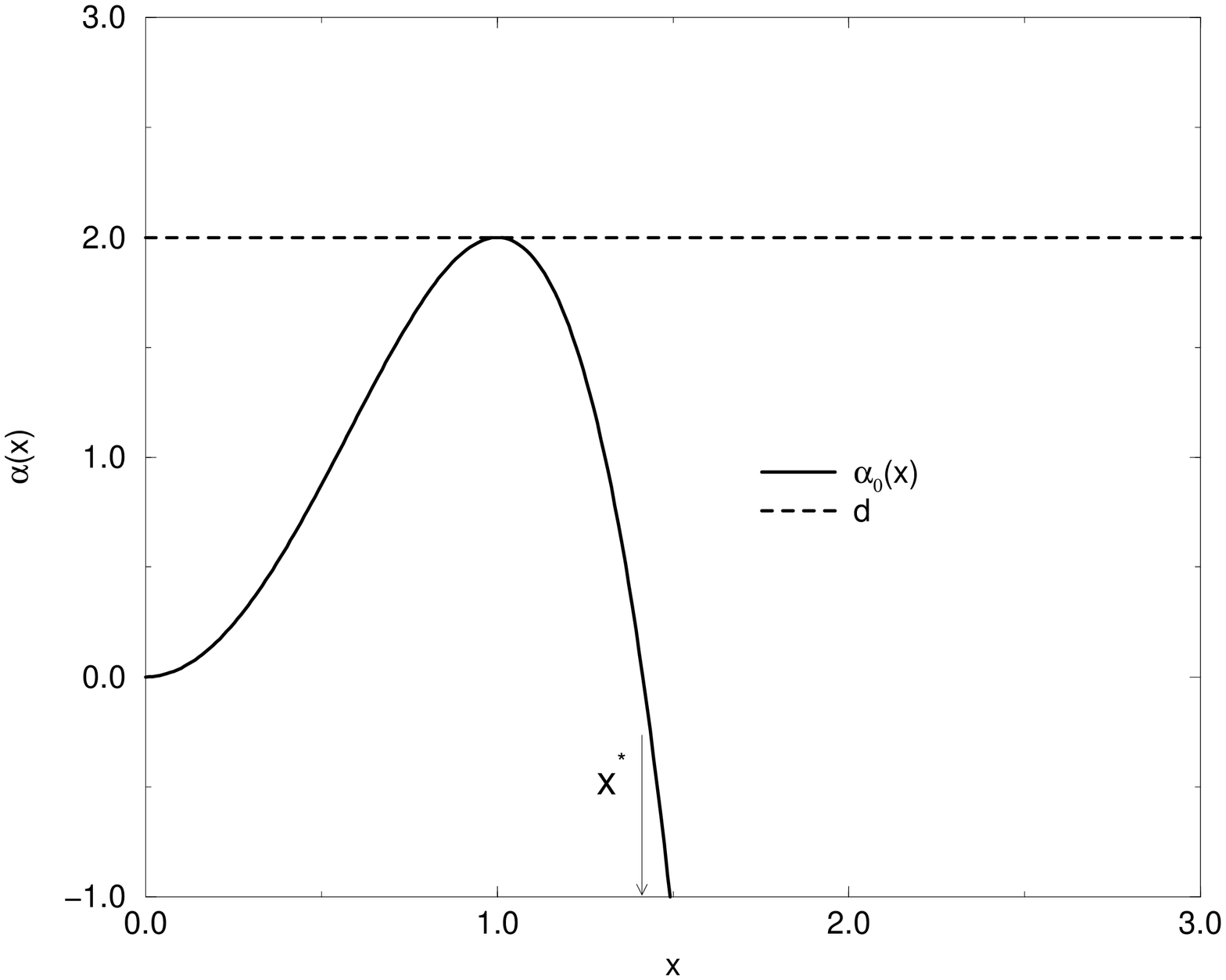,width=12cm,angle=-90}}
\caption{
Spectrum of multiscaling exponents in the large-$N$ model
with d=3, $T_F=0$ ($\alpha_0(x)$) and $T_F>0$ ($\alpha_T(x)$).
}
\label{Fig1}
\efi
\vspace{-1cm}

\bfi
\centerline{\psfig{figure=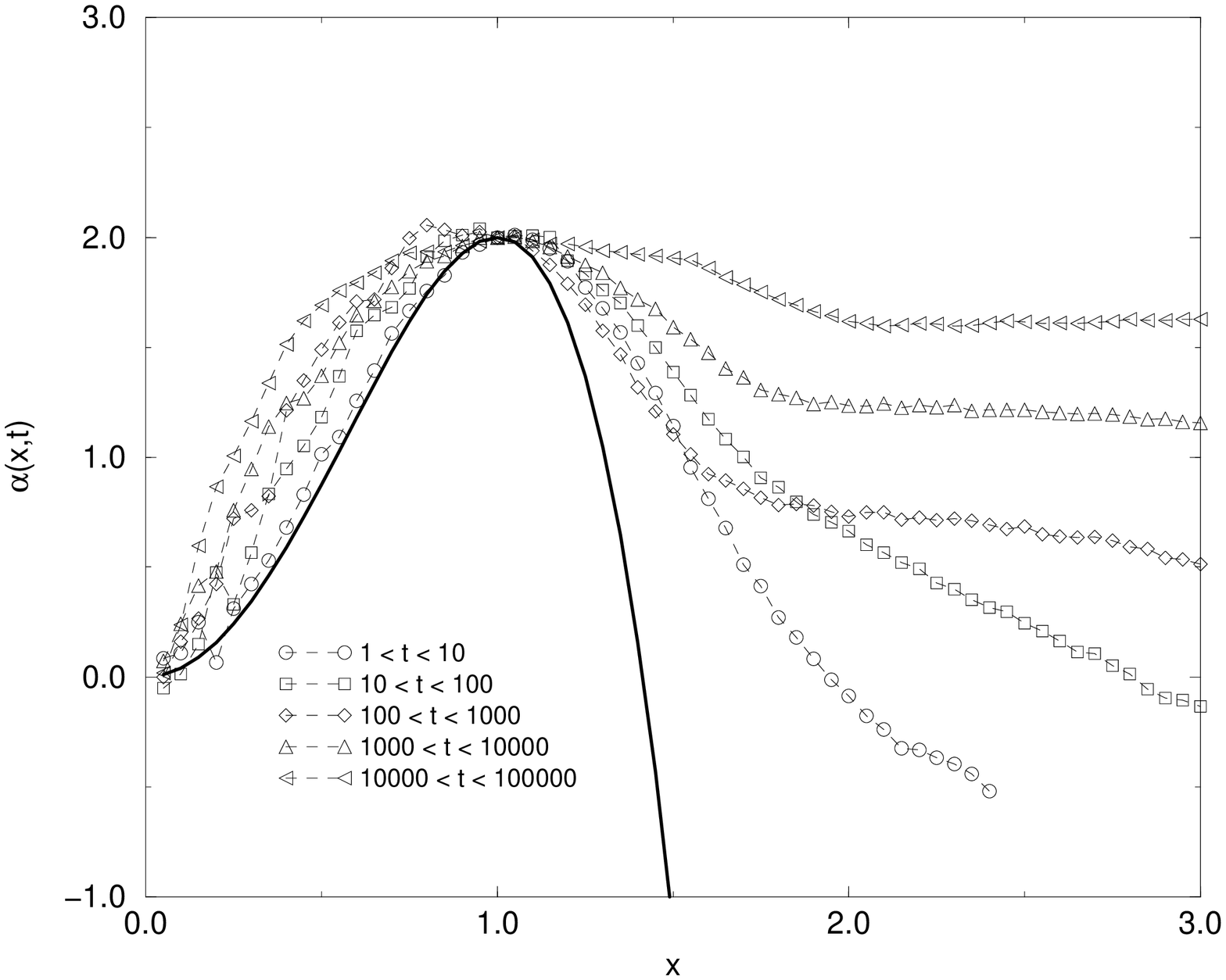,width=12cm,angle=-90}}
\caption{
Effective exponent in the quench with $T_F=0.5 T_c$ and $d=2$.
}
\label{Fig2}
\efi

\bfi
\centerline{\psfig{figure=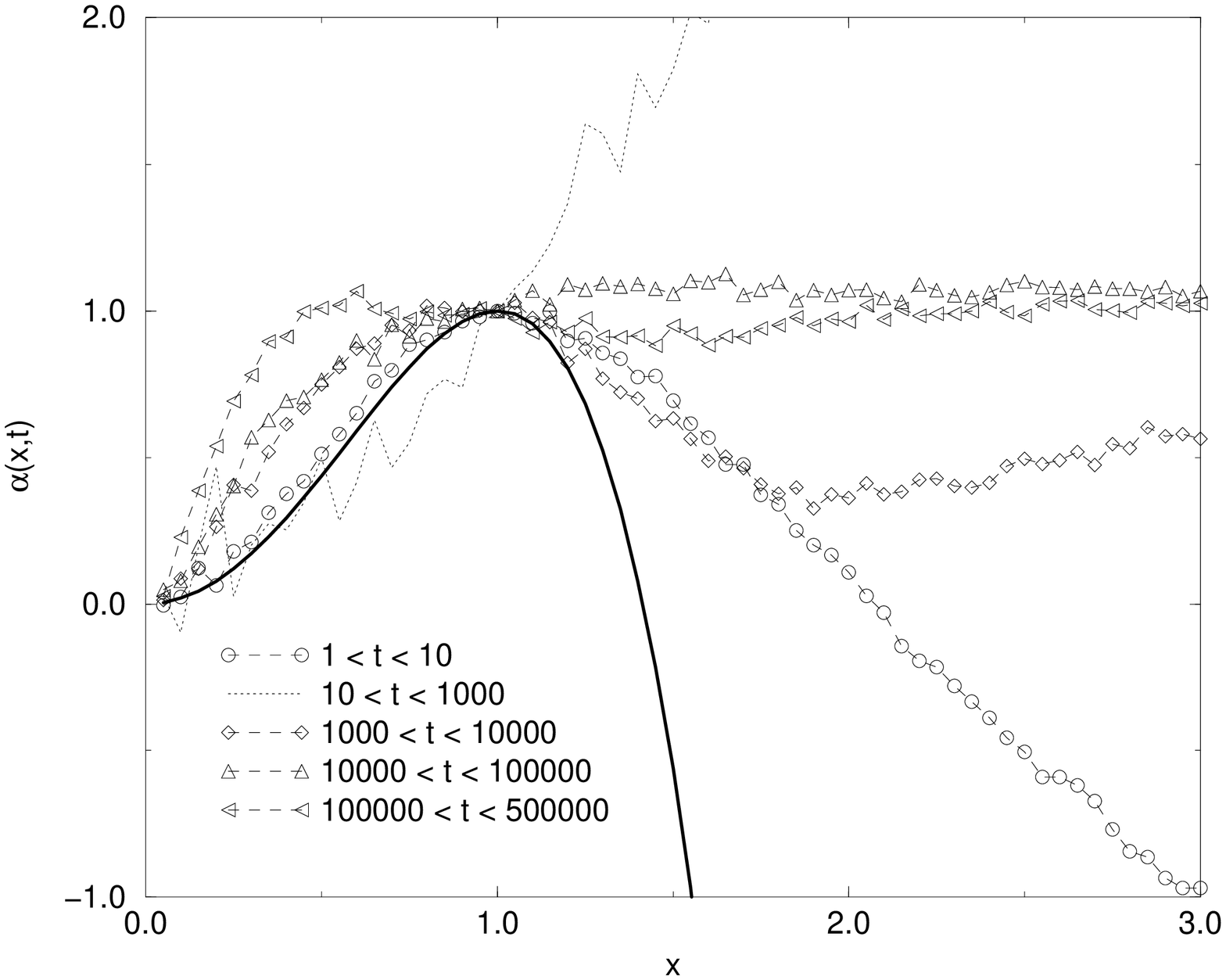,width=12cm,angle=-90}}
\caption{
Effective exponent in the quench with $T_F=0.5 J/k_{B}$ and $d=1$.
In the time interval from 10 to 1000 ${\cal L}_1=C(k_m,t)$ is almost constant.
This implies that the interval of abscissae over which $\alpha(x,t)$
is determined, as the slope of  
$\ln C(x/{\cal L}_2(t),t)$ versus $\ln {\cal L}_1(t)$, is very narrow.
As a consequence, the spectrum of effective exponents is very noisy
and no real significance can attributed to the dotted curve in the plot.
}
\label{Fig3}
\efi

\end{document}